\documentclass[12pt]{article}
\usepackage{amsfonts}
\usepackage{amsmath}
\usepackage{amssymb}
\usepackage{mathptmx}
\usepackage[dvips]{color}
\usepackage{epsfig}
\usepackage{graphicx}
\newcommand{\calA}{{\cal A}}
\newcommand{\calC}{{\cal C}}
\newcommand{\calF}{{\cal F}}
\newcommand{\calG}{{\cal G}}
\newcommand{\calL}{{\cal L}}

\newcommand{\Br}{\textrm{Br}}
\newcommand{\GeV}{{\rm GeV}}

\newcommand{\ubar}{\bar{u}}

\begin{document}
\baselineskip=16pt

\pagenumbering{arabic}

\vspace{1.0cm}

\begin{center}
{\Large\sf Lepton flavor violating muon decays in a model of
electroweak-scale right-handed neutrinos}
\\[10pt]
\vspace{.5 cm}

{Jian-Ping Bu, Yi Liao\footnote{liaoy@nankai.edu.cn}, Ji-Yuan Liu}

{\small Department of Physics, Nankai University, Tianjin 300071,
China} %

\vspace{2.0ex}

{\bf Abstract}

\end{center}

The small neutrino mass observed in neutrino oscillations is
nicely explained by the seesaw mechanism. Rich phenomenology is
generally expected if the heavy neutrinos are not much heavier
than the electroweak scale. A model with this feature built in has
been suggested recently by Hung. The model keeps the standard
gauge group but introduces chirality-flipped partners for the
fermions. In particular, a right-handed neutrino forms a weak
doublet with a charged heavy lepton, and is thus active. We
analyze the lepton flavor structure in gauge interactions. The
mixing matrices in charged currents (CC) are generally
non-unitary, and their deviation from unitarity induces flavor
changing neutral currents (FCNC). We calculate the branching
ratios for the rare decays $\mu\to e\gamma$ and $\mu\to ee\bar e$
due to the gauge interactions. Although the former is generally
smaller than the latter by three orders of magnitude, parameter
regions exist in which $\mu\to e\gamma$ is reachable in the next
generation of experiments even if the current stringent bound on
$\mu\to ee\bar e$ is taken into account. If light neutrinos
dominate for $\mu\to e\gamma$, the latter cannot set a meaningful
bound on unitarity violation in the mixing matrix of light leptons
due to significant cancelation between CC and FCNC contributions.
Instead, the role is taken over by the decay $\mu\to ee\bar e$.

\begin{flushleft}

\end{flushleft}

\newpage

Neutrino oscillation has provided the first evidence of physics
beyond the Standard Model (SM) that the neutrinos are massive,
non-degenerate and mix. The large or even maximal mixing angles
measured in solar and atmospheric neutrinos should in principle
allow the detection of lepton flavor violating (LFV) effects in
the charged lepton sector, e.g., the observation of the muon
decays, $\mu\to e\gamma$ and $\mu\to ee\bar{e}$. In simple
extensions of SM that incorporate only right-handed neutrino
singlets, this is not the case. As far as those loop-induced
processes of charged leptons are concerned, all neutrinos can be
considered as degenerate and their leading-order contribution is
thus removed by the unitarity of the mixing matrix. The tiny mass
of neutrinos diminishes the contribution further via the leptonic
GIM mechanism: all such effects are suppressed by the tiny ratio
of neutrino masses squared over those of weak gauge bosons and are
therefore not observable in the foreseeable future \cite{old}.

It would be physically more interesting if LFV effects could also
be observed beyond neutrino oscillations. The current limits on
LFV muon decays are already stringent, with the branching ratios
$\Br(\mu\to e\gamma)<1.2\times 10^{-11}$ \cite{mega99} and
$\Br(\mu\to ee\bar{e})<1.0\times 10^{-12}$ \cite{sindrum88}. The
former one will likely be pushed to $10^{-13}\sim 10^{-14}$ in the
coming years \cite{future}. Significant progress has also been
made in LFV $\tau$ decays, although the constraints are not
comparable to the muon's in the near future. An observation of
such processes will unambiguously point to non-trivial new
physics. There are indeed many alternatives for new physics that
contain new sources of lepton flavor violation. For instance, the
LFV decays could be large enough to be observable in
supersymmetric models \cite{susy}, in the extension of SM by a
Higgs triplet \cite{triplet}, and in the littlest Higgs model with
$T$-parity \cite{littlest}, to just mention a few among many
\cite{review}. For a model-independent, leading logarithmic QED
correction to the decay $\mu\to e\gamma$, see Ref.
\cite{Czarnecki:2001vf}.

The extreme smallness of neutrino mass can be understood in the
elegant seesaw mechanism \cite{seesaw}. In its standard
implementation, this is done by assuming a Dirac mass of order
charged leptons' and a huge mass of heavy neutrinos typically of
order grand unification scale. But then the heavy neutrinos that are
at the heart of new physics are beyond direct experimental
accessibility. Richer phenomenology would be possible if heavy
neutrinos had a mass not much greater than the electroweak scale so
that they could be detected at high energy colliders.

A model with the above desired feature built in has been suggested
recently by Hung \cite{Hung:2006ap}. (See also Ref.
\cite{Aranda:2007dq} for an alternative model building with
neutrinos at the electroweak scale.) The model retains the SM
gauge group albeit in a `vector-like' manner: the SM (ordinary)
fermions are augmented with mirror fermions that carry the same
charges as their SM partners but with chirality flipped. In
particular, a right-handed neutrino that is sterile in many models
now becomes a member of a weak doublet of mirror leptons. A tiny
Dirac mass for neutrinos is provided by a scalar singlet whose
vacuum expectation value is not necessarily associated with the
electroweak scale, while a Majorana mass of order the electroweak
scale is introduced by a scalar triplet. As we shall describe in
detail, this model has a rich flavor structure in weak gauge
couplings as well as in Yukawa couplings. The weak charged
couplings are generally non-unitary with or without restricting to
the subspace of light leptons, and flavor changing neutral
currents (FCNC) occur in a way that is controlled by the weak
charged couplings. It is the purpose of this work to explore their
implications for the LFV muon decays. We find that there exist
parameter regions where the decay $\mu\to e\gamma$ is accessible
in the planned experiments when the current upper bound on $\mu\to
ee\bar{e}$ is almost saturated.

We start with a brief description of the model relevant to our later
analysis; for a full account of it, see Ref. \cite{Hung:2006ap}. We
consider three generations and use slightly different notations from
the reference. The SM and mirror leptons with quantum numbers under
the gauge group $SU(2)\times U(1)_Y$ are:
\begin{eqnarray}
F_L=\left(\begin{array}{c}n_L\\f_L\end{array}\right)~({\bf
2},Y=-1),&&f_R~({\bf 1},Y=-2);\nonumber\\
F^M_R=\left(\begin{array}{c}n_R^M\\f_R^M\end{array}\right)~({\bf
2},Y=-1),&&f_L^M~({\bf 1},Y=-2);
\end{eqnarray}
where the subscripts $L,~R$ refer to chirality and the superscript
$M$ to mirror. For anomaly cancelation, the quark sector also has
mirror partners that are of no interest here. Besides the SM scalar
doublet $\Phi$, the model contains the new scalars
\begin{eqnarray}
\phi~({\bf 1},0),~\chi~({\bf 3},2),
\end{eqnarray}
plus an additional triplet $\xi~({\bf 3},0)$ that together with
$\chi$ preserves the custodial symmetry \cite{Chanowitz:1985ug} but
is irrelevant here.

The Yukawa couplings of leptons are, with the generation indices
suppressed,
\begin{eqnarray}
-\calL_{\Phi}&=&y\overline{F_L}\Phi f_R+y_M
\overline{F^M_R}\Phi f_L^M+{\rm h.c.},\nonumber\\
-\calL_\phi&=&x_F\overline{F_L}F^M_R\phi+x_f\overline{f_R}f_L^M\phi
+{\rm h.c.},\nonumber\\
-\calL_{\chi}&=&\frac{1}{2}z_M \overline{(F^M_R)^C}(i\tau^2)\chi
F^M_R+{\rm h.c.},
\end{eqnarray}
where $\psi^C=\calC\gamma^0\psi^*$, $\calC=i\gamma^0\gamma^2$, and
\begin{eqnarray}
\chi=\frac{1}{\sqrt{2}}\vec{\tau}\cdot\vec{\chi}=\frac{1}{\sqrt{2}}
\left(\begin{array}{cc}\chi^+&\sqrt{2}\chi^{++}\\\sqrt{2}\chi^0&-\chi^+
\end{array}\right).
\end{eqnarray}
A potential Majorana coupling of $\chi$ to $F_L$ is forbidden by
imposing an appropriate $U(1)$ symmetry \cite{Hung:2006ap}. Suppose
the vacuum expectation values have the structure:
\begin{eqnarray}
\langle\Phi\rangle=\frac{v_2}{\sqrt{2}}\left(\begin{array}{c}0\\1\end{array}\right),
~\langle\phi\rangle=v_1,
~\langle\chi\rangle=v_3\left(\begin{array}{cc}0&0\\1&0\end{array}\right),
\end{eqnarray}
where $v_{2,3}$ contribute to the masses of weak gauge bosons and
are naturally of order the electroweak scale while $v_1$ is not
necessarily related to it. In the basis of $f,~f^M$, the charged
lepton mass terms are
\begin{eqnarray}
-\calL^f_{\rm m}&=&\left(\overline{f_L},\overline{f_L^M}\right)
m_f\left(\begin{array}{c}f_R\\f_R^M\end{array}\right)+{\rm
h.c.},\nonumber\\
m_f&=&\left(\begin{array}{cc}\displaystyle\frac{v_2}{\sqrt{2}}y& v_1x_F\\
v_1x^{\dagger}_f&\displaystyle\frac{v_2}{\sqrt{2}}y^{\dagger}_M%
\end{array}\right),
\end{eqnarray}
while the neutrino mass terms are
\begin{eqnarray}
-\calL^n_{\rm m}&=&
\frac{1}{2}\left(\overline{n_L},\overline{(n_R^M)^C}\right)m_n
\left(\begin{array}{c}n_L^C\\n_R^M\end{array}\right)+{\rm
h.c.},\nonumber\\
m_n&=&\left(\begin{array}{cc}0&v_1x_F\\v_1x_F^T&v_3z_M\end{array}\right).
\end{eqnarray}
The seesaw mechanism operates for a Majorana mass of order the
electroweak scale and a Dirac mass proportional to $v_1$ that can be
chosen small. This relaxes in some sense the tension in ordinary
seesaw models between the generation of a light neutrino mass and
the observability of heavy neutrinos at colliders
\cite{Hung:2006ap}.

The lepton mass matrices are diagonalized by unitary transformations
($a=L,~R$):
\begin{eqnarray}
&&\left(\begin{array}{c}f\\f^M\end{array}\right)_a=X_a\ell_a,
~~~X_L^{\dagger}m_fX_R=m_\ell={\rm diag}(m_\alpha),\nonumber\\
&&\left(\begin{array}{c}n_L^C\\n_R^M\end{array}\right)=Y\nu_R,
~~~Y^Tm_nY=m_\nu={\rm diag}(m_i),
\end{eqnarray}
where $\alpha=e,\mu,\tau,\dots$ denotes the mass eigenstates of the
charged leptons and $i=1,2,3,\dots$ those of the neutrinos with the
first (last) three being light (heavy). The neutrinos are of
Majorana-type, $\nu=\nu_R+\nu_L$ with $\nu_L=\nu_R^C$. There is a
constraint on their masses from the zero texture,
$\displaystyle\sum_{k=1}^6m_kY_{ik}Y_{jk}=0$, for $i,j=1,2,3$.

The above diagonalizing matrices will enter the gauge (and Yukawa)
interactions of leptons. Some algebra yields,
\begin{eqnarray}
&&\calL_g=g_2\left(j^{+\mu}_WW_{\mu}^++j^{-\mu}_WW_{\mu}^-+J^\mu_ZZ_\mu\right)
+eJ^\mu_{\rm em}A_\mu,
\end{eqnarray}
where the currents are ($P_{L,R}=(1\mp\gamma_5)/2$)
\begin{eqnarray}
\sqrt{2}j^{+\mu}_W&=&
\bar\nu\gamma^\mu\left(V_LP_L+V_RP_R\right)\ell,
\nonumber\\
c_WJ^\mu_Z&=& \frac{1}{2}\overline{\nu}\gamma^\mu\left(
V_LV_L^\dagger P_L+V_RV_R^\dagger P_R\right)\nu
\nonumber\\%
&& -\frac{1}{2}\bar\ell\gamma^\mu\left(V_L^\dagger V_L P_L+
V_R^\dagger V_RP_R\right)\ell+s_W^2\bar\ell\gamma^\mu\ell,
\nonumber\\
J^\mu_{\rm em}&=&-\bar\ell\gamma^\mu\ell,
\end{eqnarray}
and $c_W=\cos\theta_W,~s_W=\sin\theta_W$ with $\theta_W$ being the
Weinberg angle. To relate the matrices $V_L,~V_R$ to $X_a,~Y$, it is
convenient to decompose the latter into the up and down $3\times 6$
blocks,
\begin{eqnarray}
X_a=\left(\begin{array}{c}X_a^{\rm u}\\X_a^{\rm
d}\end{array}\right), ~~~Y=\left(\begin{array}{c}Y^{\rm u}\\Y^{\rm
d}\end{array}\right),
\end{eqnarray}
then
\begin{eqnarray}
V_L=Y^{{\rm u}T}X_L^{\rm u},~V_R=Y^{{\rm d}\dagger}X_R^{\rm d},
\end{eqnarray}
with $V_L^TV_R=0$. These matrices are generally non-unitary and
the deviation from unitarity induces FCNC in both sectors of
neutrinos and charged leptons:
\begin{eqnarray}
&&V_LV_L^\dagger=Y^{{\rm u}T}Y^{{\rm u}*},~ V_RV_R^\dagger=Y^{{\rm
d}\dagger}Y^{\rm d};\nonumber\\
&&V_L^\dagger V_L=X_L^{{\rm u}\dagger}X_L^{\rm u},~V_R^\dagger
V_R=X_R^{{\rm d}\dagger}X_R^{\rm d}.
\end{eqnarray}
$X_L^{\rm d},~X_R^{\rm u}$ do not enter the charged currents (CC)
since $f_L^M,~f_R$ are $SU(2)$ singlets. Although $f_L^M,~f_R$ carry
$U(1)_Y$ charges, their electromagnetic currents are vector-like and
their neutral currents (NC) are also vector-like when combined with
those of $f_R^M,~f_L$ so that $X_L^{\rm d},~X_R^{\rm u}$ do not
enter these currents either.

\begin{figure}
\includegraphics[width=6cm]{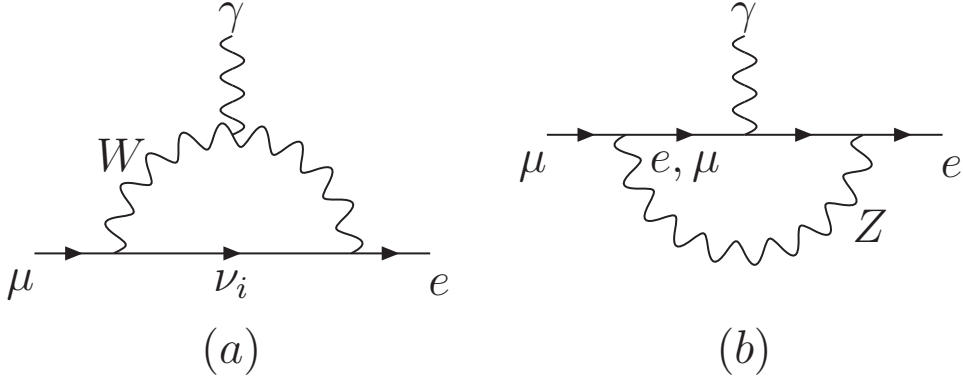}
\caption{Diagrams contributing to $\mu\to e\gamma$.} %
\label{fig1}
\end{figure}

The gauge interactions displayed above will induce LFV processes at
both tree and loop levels. The leading contribution to the decay
$\mu\to e\gamma$ occurs at one loop as shown in Fig. \ref{fig1}. The
two diagrams corresponding to CC and FCNC gauge interactions give
the following on-shell amplitudes:
\begin{eqnarray}
\calA_W&=&\frac{e}{(4\pi)^2}\sqrt{2}G_Fq^\beta\epsilon^{\alpha *}
\nonumber\\
&\times&\ubar_ei\sigma_{\alpha\beta}\left[m_\mu(V_1P_R+V_2P_L)\calF(r_i)
+m_i(V_3P_L+V_4P_R)\calG(r_i)\right]u_\mu,\nonumber\\
\calA_Z&=&\frac{e}{(4\pi)^2}\sqrt{2}G_Fq^\beta\epsilon^{\alpha *}
\nonumber\\
&\times&\ubar_ei\sigma_{\alpha\beta}
m_\mu\frac{2}{3}\left[-2(1+s^2_W)V_1P_R+(3-2s^2_W)V_2P_L\right]u_\mu,
\end{eqnarray}
where $\epsilon$ and $q$ are respectively the polarization and
momentum of the photon, and $u_{e,\mu}$ the lepton spinors. The
ratio $r_i=m_i^2/m^2_W$, and the mixing matrix elements are
\begin{eqnarray}
&&V_1=(V^\dagger_L)_{ei}(V_L)_{i\mu},~
V_2=(V^\dagger_R)_{ei}(V_R)_{i\mu},\nonumber\\
&&V_3=(V^\dagger_R)_{ei}(V_L)_{i\mu},~
V_4=(V^\dagger_L)_{ei}(V_R)_{i\mu}.
\end{eqnarray}
The summation over the neutrino index $i$ is understood in both
amplitudes. The loop functions are found to be
\begin{eqnarray}
\calF(r)&=&
\frac{1}{6(1-r)^4}\left[10-43r+78r^2-49r^3+4r^4+18r^3\ln r\right],\nonumber\\
\calG(r)&=&\frac{1}{(1-r)^3}\left[-4+15r-12r^2+r^3+6r^2\ln r\right].
\end{eqnarray}
We have taken $m_e=0$ and kept $m_\mu$ only until its linear term
that is required for chirality flip. This is a good approximation
even for the $\tau$ decays, $\tau\to e\gamma,~\mu\gamma$. In the $Z$
diagram we have ignored smaller contributions from other charged
leptons and small corrections to the diagonal $Z$ vertex so that we
stay at the same precision level as the $W$ diagram.

An interesting technical point is in order. It is simplest to work
in unitarity gauge. For the $Z$ diagram, this is all right both
because the would-be Goldstone boson contributes at a higher order
in the lepton masses than kept in the above and because the diagram
is convergent enough for the relevant Lorentz structure. But this is
not automatically true with the $W$ diagram which is more
ultraviolet divergent due to the triple gauge coupling. There is no
guarantee in this case that the order of removing the ultraviolet
regulator commutes with that of taking the unitarity gauge limit. As
a matter of fact, although the diagram is convergent in both
unitarity and $R_\xi$ gauges, there is a finite difference in the
terms linear in the lepton masses between the results obtained in
the two gauges. This caveat is restricted to the mentioned terms
because terms of a higher order are convergent enough to allow the
free interchange of taking the limits. In the conventional case of
unitary, pure left-handed couplings, the linear terms are killed by
the unitarity of $V_L$ so that an identical result can be reached in
either gauge \cite{book}. This is no more the case here. Considering
this, we have replaced the terms linear in either $m_\mu$ or $m_i$
obtained in unitarity gauge by those obtained in $R_\xi$ gauge whose
$\xi$ dependence is canceled as expected.

The above amplitude involves several neutrino masses and many
mixing matrix elements. In our later numerical analysis, we shall
make some approximations. First, the light neutrinos can be safely
treated as massless. Then, $\calF\to\frac{5}{3}$, and the $\calG$
term multiplied by $m_i$ can be ignored. In simple extensions of
SM, the leading term of $\calF$ is removed by the unitarity of the
CC mixing matrix of light leptons while the $\calG$ term does not
appear, leaving behind a significantly GIM suppressed term that is
not observable \cite{old}. This is not the case in the type of
models considered here. From the phenomenological point of view,
neutrino oscillation experiments that are the main source of the
lepton mixing matrix so far, are not yet precise enough to test
its unitarity. Instead, it is exactly the lepton flavor changing
transitions studied here that provide the most stringent
constraint on the unitarity. Second, we assume that the heavy
neutrinos are almost degenerate. We checked that the leading terms
of $\calF(r)$ and $\calG(r)$ in the limit $r\to\infty$ deviate
significantly from the exact values for $m_i$ of order the
electroweak scale or slightly higher. We shall thus retain their
exact forms for numerical analysis. As a bonus of the
approximations, the amplitude depends on the products of matrix
elements summed over light and heavy neutrinos respectively:
\begin{eqnarray}
&&V^l_1=\sum_{i=1}^3(V^\dagger_L)_{ei}(V_L)_{i\mu},~
V^l_2=\sum_{i=1}^3(V^\dagger_R)_{ei}(V_R)_{i\mu},\nonumber\\
&&V^l_3=\sum_{i=1}^3(V^\dagger_R)_{ei}(V_L)_{i\mu},~
V^l_4=\sum_{i=1}^3(V^\dagger_L)_{ei}(V_R)_{i\mu},
\end{eqnarray}
and similarly for $V^h_{1,2,3,4}$ with $i$ summed over $4,5,6$.

The branching ratio is then
\begin{eqnarray}
\Br(\mu\to e\gamma)=
\frac{3\alpha}{8\pi}\left(|h_L|^2+|h_R|^2\right),
\end{eqnarray}
where, denoting the common heavy neutrino mass as $m_h$ and
$r_h=m^2_h/m^2_W$,
\begin{eqnarray}
h_L&=&\frac{5}{3}V^l_2+V^h_2\calF(r_h)
+\frac{m_W}{m_\mu}V^h_3\sqrt{r_h}\calG(r_h)
+\frac{2}{3}(3-2s^2_W)(V^l_2+V^h_2),\nonumber\\
h_R&=&\frac{5}{3}V^l_1+V^h_1\calF(r_h)
+\frac{m_W}{m_\mu}V^h_4\sqrt{r_h}\calG(r_h)
-\frac{4}{3}(1+s^2_W)(V^l_1+V^h_1).
\end{eqnarray}
Note in passing that the heavy neutrinos do not necessarily
decouple in the heavy mass limit. For $r\to\infty$,
$\calF(r)\to\frac{2}{3}$ and $\calG(r)\to -1$. The explicit factor
$m_h$ appearing in front of $\calG(r_h)$ is actually canceled by
$m^{-1}_h$ coming from $V^h_{3,4}$, since the latter are
proportional to $v_1x_F/m_h$ with $v_1x_F$ being independent of
$m_h$ to good precision. The contribution to the same process from
the heavy charged leptons-$\phi$ loop has recently been considered
in Ref. \cite{Hung:2007ez} in the heavy lepton limit. The singlet
scalar $\phi$ has been assumed not to mix with other scalars. Note
that even with this simplifying assumption the coupling matrices
involved in the two types of contributions cannot be mutually
obtained. In particular, the neutrino diagonalizing matrix $Y$
does not enter into the $\phi$ diagram.

Now we turn to the decay $\mu\to ee\bar{e}$ whose leading term
occurs at the tree level via FCNC. There are two diagrams due to
identical fermions appearing in the final state. Once again, we
ignore the small correction to the diagonal $Ze\bar{e}$ vertex in
SM. Taking into account a factor of $\frac{1}{2}$ in the phase
space, the branching ratio is
\begin{eqnarray}
\Br(\mu\to
ee\bar{e})&=&\frac{1}{2}|V_1^l+V_1^h|^2\left[(1-2s^2_W)^2+2s^4_W\right]
\nonumber\\
&+&\frac{1}{4}|V_2^l+V_2^h|^2\left[(1-2s^2_W)^2+8s^4_W\right].
\end{eqnarray}

The two branching ratios involve the following unknown parameters:
the six complex matrix elements in the form of $V_{1,2}^l$,
$V_{1,2,3,4}^h$ plus one heavy neutrino mass $m_h$. Roughly
speaking, for all matrix elements of similar order and $m_h$
deviating not much from $m_W$, we have $\Br(\mu\to
e\gamma)/\Br(\mu\to ee\bar{e})\sim\frac{\alpha}{\pi}\sim 2\times
10^{-3}$. One cannot get better quantitative feel of the effects
without making some further simplifications. To demonstrate the
physical relevance of our results, we choose to present our
numerical results by sampling $m_h$ and the matrix elements in
certain ranges. We consider the following scenarios for the purpose
of illustration. For the standard input parameters, we use
$\alpha=1/137.04,~m_W=80.2~\GeV,~m_\mu=0.1056~\GeV,~s^2_W=0.23$.

\begin{figure}
\includegraphics[width=15cm]{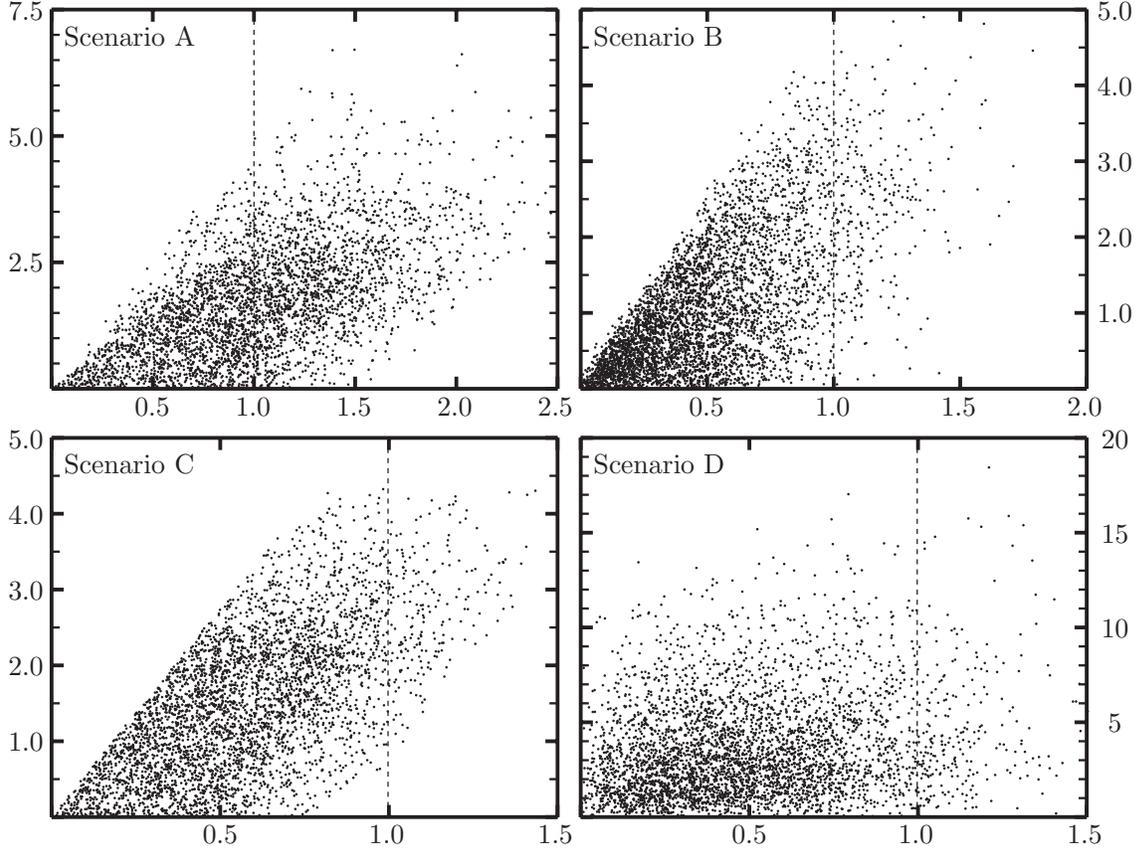}
\caption{Sampled points for $\Br(\mu\to ee\bar{e})$ (horizontal, in
units of $10^{-12}$) and $\Br(\mu\to e\gamma)$ (vertical, in units
of $10^{-14}$) for the four scenarios described in the text. The
dashed vertical line shows the current upper bound on $\Br(\mu\to
ee\bar{e})$.} %
\label{fig2}
\end{figure}

We find an algebraically simple case after some inspection.
Suppose the upper-right $3\times 3$ block of $Y$ is real. In this
scenario A, our special neutrino spectrum (three almost massless
plus three almost degenerate and heavy) implies that the two
off-diagonal $3\times 3$ blocks of $Y$ vanish, the lower-right
block is trivially identity and the upper-left one is unitary.
Then, $V_1^l=(x_L^\dagger x_L)_{e\mu}$, $V^h_2 =-(x_R^\dagger
x_R)_{e\mu}$ while all others vanish, where $x_{L,R}$ are the
upper-left $3\times 3$ blocks of $X_{L,R}$ respectively. Since we
have no idea of their magnitudes, we sample randomly the real and
imaginary parts of $V_1^l,~V_2^h$ between $-2\times 10^{-6}$ and
$+2\times 10^{-6}$, keeping an eye on the current upper bound on
$\Br(\mu\to ee\bar e)$. For the heavy neutrino mass we choose
$m_h=50,100$ up to $1000$ GeV. The combined result is shown in
Fig. \ref{fig2}. For $\Br(\mu\to ee\bar e)<10^{-12}$, most points
drop in the region where $\Br(\mu\to e\gamma)$ is at the edge of
precision available in the next generation of experiments, $\sim
10^{-14}$.

In scenario B, we sample the real and imaginary parts of
$V^l_{1,2},~V^h_{1,2}$ in the range $[-10^{-6},10^{-6}]$ while
keeping $V^h_3=V^h_4=0$ and assuming the value of $m_h$ as in
scenario A. The matrix elements are chosen smaller than in scenario
A in order that most points would not break the current bound on
$\Br(\mu\to ee\bar e)$. The terms from the four elements tend to
interfere constructively so that $\Br(\mu\to e\gamma)$ is slightly
larger than in scenario A.

We assume in scenario C that only the contribution of light
neutrinos is important while that of heavy ones is suppressed for
some reason. The real and imaginary parts of $V^l_1,~V^l_2$ run
randomly in the range from $-1.5\times 10^{-6}$ to $+1.5\times
10^{-6}$, and the result is independent of $m_h$. We find that
$\Br(\mu\to e\gamma)\lesssim {\rm a~few~}\times 10^{-14}$ for
$\Br(\mu\to 3e)<10^{-12}$ in most regions of the parameter space.
Actually, this scenario can be better treated analytically. The
branching ratios are
\begin{eqnarray}
\Br(\mu\to e\gamma)&\approx&
10^{-4}\left[0.0064|V_1^l|^2+102|V_2^l|^2\right],\nonumber\\
\Br(\mu\to ee\bar{e})&\approx&0.20|V_1^l|^2+0.18|V_2^l|^2.
\end{eqnarray}
The very small coefficient, $(1-4s_W^2)^2=0.0064$, of $|V_1^l|^2$
in $\Br(\mu\to e\gamma)$ arises from the destructive interference
between the $W$ and $Z$ graphs. If the $W$ graph were only
present, the coefficient would be $25$. This is indeed the case in
the models where FCNC does not appear in the charged lepton
sector; and the light neutrino contribution to $\mu\to e\gamma$
via pure left-handed CC gauge interactions (i.e., $V_2^l=0$) has
been employed in Ref. \cite{Antusch:2006vwa} to set a stringent
upper bound on unitarity violation in the mixing matrix of light
leptons (i.e., $|V_1^l|^2$). However, for the type of new physics
as discussed here in which FCNC occurs in both sectors of leptons,
we can no longer utilize the decay to set a useful bound on
$|V_1^l|^2$ as its effect has been diminished by a factor of
$25/0.0064\sim 3900$. In this case, the decay $\mu\to ee\bar e$
studied here sets a much more stringent bound, $|V_1^l|^2<5\times
10^{-12}$. This means that we can ignore $V_1^l$ for $\mu\to
e\gamma$. Using again the bound from $\mu\to ee\bar e$, this
implies in turn an upper bound on $\mu\to e\gamma$ in this
scenario:
\begin{eqnarray}
\Br(\mu\to e\gamma)\approx 10^{-2}|V_2^l|^2<5.7\times 10^{-13}.
\end{eqnarray}
The best one can have is to saturate the above bound on $\mu\to
e\gamma$ while sitting at the current experimental bound on $\mu\to
ee\bar e$. It is impossible in particular to approach a branching
ratio of $10^{-12}$ for both decays simultaneously.

To get some feel on the mixed effect between left- and right-handed
CC currents involving light charged leptons and heavy neutrinos, we
consider scenario D. The real and imaginary parts of
$V^{l,h}_1,~V^{l,h}_2$ are allowed to run randomly in the range from
$-10^{-6}$ to $+10^{-6}$ while the range of $V^h_{3,4}$ is smaller
by a factor of $10^{-3}$. The latter two are likely smaller than
others since they involve the Dirac neutrino mass term proportional
to $v_1x_F$ where $v_1$ is small \cite{Hung:2006ap}. It seems
difficult to get an exact handle of the orders of magnitude on the
involved matrix elements since the heavy charged lepton masses also
set in through the diagonalizing matrices $X_{L,R}$. We thus choose
to illustrate our results by assuming a value for $m_h$ from $50$
GeV to $500$ GeV at a step of $50$ GeV when sampling $V$'s. We do
not assume a larger value for it to avoid amplifying artificially
the heavy neutrino term because as we mentioned earlier $V_{3,4}^h$
is proportional to $m^{-1}_h$ to good precision. We find that
$\Br(\mu\to e\gamma)$ can reach the level of $10^{-13}$ for
$\Br(\mu\to 3e)<10^{-12}$.

The small neutrino mass is naturally explained by the seesaw
mechanism. Physics would be phenomenologically more interesting if
heavy neutrinos have a mass close to the electroweak scale. In
that case, they would be directly accessible at high energy
colliders. On the other hand, the large leptonic mixing observed
in neutrino oscillations does not imply large lepton flavor
violation in the charged lepton sector if the neutrino mass is
incorporated in a trivial manner. An observation of LFV charged
lepton decays would thus point to non-trivial new physics related
to the origin of neutrino mass. This is encouraged especially by
experimental advances expected in the near future. Motivated by
this observation, we have studied the rare decays $\mu\to
e\gamma,ee\bar e$ in a model suggested recently in which
non-trivial new physics does appear with heavy neutrinos at the
electroweak scale. Although $\Br(\mu\to e\gamma)$ is generally
smaller than $\Br(\mu\to ee\bar e)$ by three orders of magnitude,
there exists a significant portion of the parameter space in which
$\Br(\mu\to e\gamma)$ reaches or is within the sensitivity
available in the new generation of experiments without breaking
the current bound on $\mu\to ee\bar{e}$. But it is generally
impossible to reach the level of $10^{-12}$ for both decays
simultaneously. When the direct contribution from heavy neutrinos
enters, it is difficult to make a definite quantitative prediction
due to too many free parameters. But if for some reason the effect
of heavy neutrinos is strongly suppressed compared to light
neutrinos, the situation becomes transparent. Due to the
destructive interference between the CC and FCNC interactions, the
decay $\mu\to e\gamma$ is insensitive to the unitarity violation
in the sector of light leptons. Instead, the other one $\mu\to
ee\bar{e}$ proceeding through tree level FCNC can set a stringent
bound on it. In this scenario, the best one can expect for the
decays is $\Br(\mu\to e\gamma)\sim 5\times 10^{-13}$ and
$\Br(\mu\to ee\bar e)\sim 10^{-12}$.

\vspace{0.5cm}
\noindent %
{\bf Acknowledgement} This work is supported in part by
the grants NCET-06-0211 and NSFC-10775074.


\end{document}